\documentclass[
reprint,
superscriptaddress,
nofootinbib,
amsmath,amssymb,
aps,prd
]{revtex4-2}
\usepackage{xcolor}
\usepackage{graphicx}
\usepackage{dcolumn}
\usepackage{bm}
\usepackage[colorlinks=true,citecolor=blue,urlcolor=blue]{hyperref}
\usepackage{amsmath}
\usepackage{subcaption}
\usepackage{ragged2e}

\begin{document}

\preprint{APS/123-QED}

\title{Spin relaxation of heavy quarks from momentum diffusion in the quark-gluon plasma}

\author{Ankit Kumar}
\email{kumar.ankit@iitgn.ac.in}
\affiliation{Indian Institute of Technology Gandhinagar, Gandhinagar-382355, Gujarat, India}

\author{Sourav Dey}
\email{sourav.d@tifr.res.in}
\affiliation{Indian Institute of Technology Gandhinagar, Gandhinagar-382355, Gujarat, India}
\affiliation{Department of Theoretical Physics, Tata Institute of Fundamental Research, Homi Bhabha Road, Mumbai, 400005, India}

\author{Vinod Chandra}
\email{vchandra@iitgn.ac.in}
\affiliation{Indian Institute of Technology Gandhinagar, Gandhinagar-382355, Gujarat, India}

\author{Amaresh Jaiswal}
\email{a.jaiswal@niser.ac.in}
\affiliation{School of Physical Sciences, National Institute of Science Education and Research, An OCC of Homi Bhabha National Institute, Jatni-752050, India}

\date{\today}

\begin{abstract}

Heavy-quark spin dynamics in the quark-gluon plasma provides a potential probe of the interaction between heavy quarks and the medium. We develop a stochastic framework in which the random momentum kicks responsible for heavy-quark diffusion also induce spin rotation through Thomas precession. Starting from the Thomas-Bargmann-Michel-Telegdi equation and a Langevin equation for heavy-quark momentum, we derive a novel coupled spin-momentum Fokker-Planck equation. In the heavy-quark limit, where momentum equilibrates faster than spin, the momentum degrees of freedom can be integrated out, yielding a rotational Fokker-Planck equation for the spin orientation. This establishes a direct relation between the Thomas precession contribution to the spin-relaxation time $\tau_{s}$ and the heavy-quark momentum-diffusion coefficient $\kappa$. In the non-relativistic limit, we find $\tau_s=2\tau_p(m_Q/T)^2$, with $\tau_p$ being the momentum relaxation time. Using recent determinations of $\kappa$ from lattice QCD, strong-coupling estimates, $T$-matrix calculations, and heavy-flavor phenomenology, we obtain the corresponding Thomas precession contribution to spin relaxation times for charm and bottom quarks. Relativistic kinematics modify the result by only a modest amount over the temperature range relevant for heavy-ion collisions, while the uncertainty is dominated by that of $\kappa$. We emphasize that the result represents the Thomas precession contribution to heavy-quark spin relaxation; direct coupling to the chromomagnetic field provides an additional channel.

\end{abstract}


\maketitle


\section{Introduction}

In non-central heavy-ion collisions (HIC), ultra-strong magnetic fields of the order of $m_{\pi}^2\approx 10^{14}\,\,\text{T}$ are generated due to the fast-moving charged spectator particles \cite{Kharzeev:2007jp, Skokov:2009qp, Voronyuk:2011jd, Bloczynski:2012en}. Although the electromagnetic fields are extremely strong initially, they decay rapidly and are substantially reduced by the time the QGP is formed~\cite{Tuchin:2013ie, Huang:2022qdn, Jiang:2022uoe, Huang:2017tsq, McLerran:2013hla}. Heavy quarks (Charm and Bottom), which are produced in the initial hard scattering processes, experience these magnetic fields and their spins get polarized in the direction of the fields \cite{Dey:2025ail, Liu:2024hii, Li:2025ipk}. When these polarized heavy quarks (HQs) pass through the QGP medium formed $\sim 1\,\,\text{fm/c}$ after the initial scattering, they experience random kicks from the particles of the QGP medium. Due to the heavy mass of the HQ ($m_Q\gg T$), these kicks only slightly affect the momentum and spin of the HQ each time these particles collide, due to which the momentum relaxation time of the HQs is $\sim m_Q/T$ times longer than the relaxation time of the light particles \cite{PhysRevD.73.014508, Moore:2004tg}. Therefore, the dynamical evolution of the HQ can be described as the Brownian motion within the QGP medium \cite{Svetitsky:1987gq, Moore:2004tg, vanHees:2005wb, vanHees:2007me, Akamatsu:2008ge, Das:2010tj, rapp2010heavy}. 

Since the HQs are produced in the initial phases of the HIC, they experience the entire evolution of the produced medium, and hence, are excellent probes to study the properties of the produced QGP medium \cite{ALICE:2020iug, STAR:2017kkh}. The evolution of polarization of the HQ, and its transfer into the open heavy flavor hadron observables, provides a sensitive probe into early time dynamics of the magnetic field and the initial geometry of the fireball \cite{Dey:2025ail, Jaiswal:2026ixt, Jaiswal:2026juz, Acharya2025, ALICE:2022byg, Dey:2026epy}. Although the momentum diffusion of the HQ in QGP has been studied well \cite{Svetitsky:1987gq, Moore:2004tg, vanHees:2005wb, vanHees:2007me, Akamatsu:2008ge, Das:2010tj, PhysRevD.85.014510, Kaczmarek:2014jga, Ding:2012iy, Herzog:2006gh}, the spin diffusion has not been explored much \cite{Dey:2025ail, Jaiswal:2026juz, Hongo:2022izs, Liu:2024hii, Li:2025ipk, Li:2025ugb}. 

In the context of atomic and solid-state physics, it is well known that the correct magnitude of the spin-orbit interaction was achieved semiclassically only after incorporating the relativistic effect of Thomas precession~\cite{Thomas:1926dy}. This effect leads to spin precession due to the non-commutativity of Lorentz boosts in different directions, in particular when the particle moves in a non-inertial frame. Here we investigate such an effect on the spin degrees of freedom of a heavy probe arising from the random change of frame due to stochastic momentum kicks from the bath degrees of freedom. This momentum stochasticity also translates to the spin degrees of freedom via a random Thomas precession effect, leading to a spin diffusion effect, which is the main finding of this study. Moreover, one can expect that the ratio of the spin diffusion coefficient $\mathcal{D}$, which characterizes the time scale $\tau_{s}=\mathcal{D}^{-1}$ over which the initial polarization of the heavy probe dissolves into the medium, to the momentum drag coefficient should be proportional to the dimensionless quantity $\left(T/m_{Q}\right)^{n}$. Here, $n$ differs from zero because, in the heavy-mass limit of the HQ ($m_Q\rightarrow\infty$), the system possesses spin-SU(2) symmetry, which freezes the spin degrees of freedom, whereas momentum diffusion can still take place. Only a higher-order mass correction, relative to the momentum case, can render the spin d.o.f. dynamical. Moreover, such a ratio can also be useful for constraining phenomenological observables in the context of both spin and momentum diffusion.

According to Ehrenfest's theorem, the expectation values of quantum operators can, under appropriate semiclassical conditions, evolve according to classical equations of motion. For a HQ spin interacting with the fluctuating color fields of the QGP, such a description is appropriate when the characteristic spatial scale of the field fluctuations is much larger than the extent of the HQ wave packet. The stochastic momentum noise experienced by the HQ originates from fluctuations of the color-electric fields in the medium, whose characteristic spatial scale is set by the inverse Debye mass, $m_D^{-1}$. In the QGP, $m_D\sim gT$, whereas the typical de Broglie wavelength of a heavy quark near thermal equilibrium is $\lambda_{\rm dB}\sim1/\sqrt{m_QT}$. Thus, in the heavy-quark limit $m_Q\gg T$, one has parametrically $\lambda_{\rm dB}\ll m_D^{-1}$, supporting a semiclassical treatment in which the HQ spin can be separated from the spatial variation of the medium fields. Furthermore, at leading order, the heavy-quark polarization obtained from a spin-density-matrix treatment was shown to be consistent with the classical stochastic description employed here~\cite{Jaiswal:2026juz}. These considerations motivate treating the HQ spin as a classical degree of freedom and describing its stochastic evolution in analogy with the established Langevin description of heavy-quark momentum diffusion.

In this paper, we construct a Langevin equation for the stochastic evolution of the HQ spin using the classical Thomas-Bargmann-Michel-Telegdi (Thomas-BMT) equation~\cite{PhysRevLett.2.435, leader2001spin}, following the standard framework of stochastic processes~\cite{Risken1996, Balakrishnan2021, GarciaPalacios2007}. In this formulation, the random momentum fluctuations that drive heavy-quark diffusion also induce stochastic Thomas precession of the spin, thereby providing a direct connection between momentum diffusion and spin relaxation. The spin relaxation time associated with this Thomas precession mechanism can then be obtained from the resulting stochastic dynamics. The origin of the precession is the kinematic fact that two successive non-collinear Lorentz boosts do not combine into a pure boost; instead, they generate a residual rotation of the particle's rest frame, giving rise to Thomas precession.

The random momentum kicks that the HQ receives from the medium change the direction of its velocity and, through the Thomas precession, rotate its spin. The same stochastic noise $\bm{\xi}$ is therefore responsible for the diffusion of both the momentum and the spin of the HQ. From the Langevin equations for the momentum and the spin, we derive a novel spin-momentum Fokker-Planck equation in the extended phase space of momentum and spin. Integrating over the momentum degrees of freedom, we obtain a Fokker–Planck equation for the spin that describes rotational diffusion on the spin sphere and has the same form as that derived in Ref.~\cite{Dey:2025ail}, with the spin relaxation time, $\tau_s$, determined by the heavy-quark momentum-diffusion coefficient $\kappa$.

The key advance over Ref.~\cite{Dey:2025ail} is that the spin-relaxation time is no longer treated as an independent phenomenological parameter. By deriving the spin dynamics directly from the Thomas–BMT equation coupled to the heavy-quark Langevin dynamics, we obtain a microscopic relation between the Thomas precession contribution to spin relaxation and the heavy-quark momentum-diffusion coefficient $\kappa$. This relation allows existing determinations of $\kappa$ to be translated into quantitative estimates of the corresponding Thomas precession spin-relaxation timescale. We estimate $\tau_s$ for charm quarks using $\kappa$ from lattice QCD \cite{HotQCD:2025fbd}, AdS/CFT \cite{PhysRevD.74.085012, casalderrey2014gauge, GUBSER2008175, Casalderrey-Solana:2007ahi}, the T-matrix approach \cite{Tang:2023tkm}, and transport model constraints on D-meson observables \cite{ALICE:2021rxa}, for both the non-relativistic and relativistic dispersion relations of the HQ. Since the vector polarization of the HQ decays on the time scale $\tau_s/2$, these estimates quantify how quickly the polarization is lost through this mechanism.


\section{Thomas precession-induced spin dynamics}

To describe the classical evolution of the spin $\bm{s}$ of the HQ, we begin with the Thomas-BMT equation for the rest frame spin vector $\bm{s}$ of the HQ as a function of laboratory time $t$ \cite{PhysRevLett.2.435, leader2001spin}
\begin{align}\label{BMT}
\frac{d\bm{s}}{dt}=\bm{s}\times\Bigg[\frac{g\mu_0}{\hbar s}&\Big(\bm{B}-\frac{\gamma}{\gamma+1}(\bm{v}\cdot\bm{B})\bm{v}-\bm{v}\times\bm{E}\Big) \nonumber\\
&+\frac{\gamma^2}{\gamma+1}\bm{v}\times\frac{d\bm{v}}{dt}\Bigg]
\end{align}
where $\mu_0$ is the magnetic moment ratio and $g$ is the gyro-magnetic ratio. The first term of Eq.~\eqref{BMT} corresponds to the interaction of the HQ spin with the background electromagnetic field. However, as noted above, although the electromagnetic fields are initially extremely strong, they decay rapidly, on a timescale of $\lesssim 1$ fm/$c$, and are substantially diminished by the time the QGP is formed~\cite{Huang:2017tsq}. Therefore, the term describing the interaction of the electromagnetic field with the HQ spin is strongly suppressed and can be neglected. The rate of change of the canonical spin vector $\bm{s}$ with respect to laboratory time $t$ is then determined solely by the last term in Eq.~\eqref{BMT}, which describes Thomas precession, and takes the form
\begin{align}
\frac{d\bm{s}}{dt}\approx \bm{s}\times \Big(\frac{\gamma^2}{\gamma+1}\bm{v}\times \frac{d\bm{v}}{dt}\Big).
\end{align}
The spin of the HQ can also couple directly to the colour-magnetic fields of the QGP medium \cite{Hongo:2022izs}. This coupling is not considered in the present work and is left for future. On the other hand, we consider the polarization generated during the early strong electromagnetic-field stage as the initial condition for the subsequent stochastic evolution in the QGP.


\section{Spin-momentum Fokker–Planck equation}

Using the momentum evolution equation given by
\begin{align}
\frac{d\bm{p}}{dt}&=-\eta_D\bm{p}+\bm{\xi}(t) \label{peq},
\end{align}
where $\eta_D$ and $\bm{\xi}(t)$ are the momentum drag coefficient and random momentum kicks experienced by the HQ passing through the medium, one can obtain the evolution equation for the spin as
\begin{align}\label{seq}
\frac{d \bm{s}}{dt}&=\frac{\bm{s}\times (\bm{p}\times \bm{\xi}(t))}{m_Q(E+m_Q)}.
\end{align}
Using Eqs.~\eqref{peq} and \eqref{seq}, we can derive the Fokker-Planck equation of the HQ in the extended phase space with spin and momentum. The above equations can be written in the component form as
\begin{align}
    \frac{dp_i}{dt}&=-\eta_D p_i+\xi_i(t),\label{compp}\\
    \frac{ds_i}{dt}&=\frac{\epsilon_{ijk}\epsilon_{klm}s_jp_l\xi_m}{m_Q(E+m_Q)}\label{comps}
\end{align}
where $\xi_i(t)$ represents the independent white noise with the following statistical properties \cite{Moore:2004tg, Akamatsu:2008ge}:
\begin{align}
    \langle\xi_i(t)\rangle&=0, \label{mean} \\
    \langle\xi_i(t_1)\xi_j(t_2)\rangle&=\kappa \delta_{ij}\delta(t_1-t_2), \label{variance}
\end{align}
where $\kappa$ is the momentum diffusion coefficient. Using Eqs.~\eqref{compp} and \eqref{comps}, we can identify the drag and diffusion coefficients as
\begin{align}\label{AandC}
A_I=
\begin{pmatrix}
-\eta_D\, \bm{p} \\
\bm{0}
\end{pmatrix},
&& C_{IJ}=
\begin{pmatrix}
I_{3\times3} & {0}_{3\times3} \\
\left.\dfrac{\left(p_is_j-\delta_{ij}\bm{s}\cdot\bm{p}\right)}{m_Q(E+m_Q)}\right|_{3\times3} & 0_{3\times3}
\end{pmatrix},
\end{align}
and rewrite Eqs.~\eqref{compp} and \eqref{comps} as the standard Langevin equation
\begin{align}\label{LE}
\frac{dy_I}{dt}=A_I+\sum_{J}C_{IJ}\xi_J, && I,J=1,2
\end{align}
where
\begin{align}
y_I=
\begin{pmatrix}
\bm{p}\\
\bm{s}
\end{pmatrix}
\quad\text{and}\quad
\xi_I=
\begin{pmatrix}
\bm{\xi}\\
\bm{0}
\end{pmatrix}.
\end{align}
For a Markov process described by Eq.~\eqref{LE}, the Fokker-Planck equation for the probability $P(y_I,t)$ can be written as \cite{Dey:2025ail, Dey:2026epy}
\begin{widetext}
\begin{align}
\frac{\partial P(y,t)}{\partial t}=-\sum_I \frac{\partial}{\partial y_I}\left[P(y,t)\left(A_I +\frac{\kappa}{2}\sum_{J,K}C_{JK}\frac{\partial C_{IK}}{\partial y_J}\right)\right]+\frac{\kappa}{2}\sum_{IJ}\frac{\partial^2}{\partial y_I\partial y_J}\left[P(y,t)\sum_K C_{IK}C_{JK}\right]. \label{FP_gen}
\end{align}
Using Eq.~\eqref{AandC} in the above equation, we get the Fokker-Planck equation for the momenta and spin of the HQ in the medium as
\begin{align}\label{FPps}
\frac{\partial P(\bm{p},\bm{s},t)}{\partial t}=&\,\frac{\partial}{\partial p_i} (P\,\eta_D\,p_i)+\frac{\kappa}{2m_Q^2(E+m_Q)^2} \frac{\partial}{\partial s_i} \left[P\,p^2\,s_i+P\,p_i\,(\bm{s}\cdot\bm{p})\right]+\frac{\kappa}{2} \frac{\partial^2 P}{\partial p_i\partial p_i} +\frac{\kappa}{m_Q(E+m_Q)}  \\
&\times \frac{\partial^2}{\partial p_i\partial s_j}\left[ P\, p_j\, s_i - P\,\delta_{ij}\,(\bm{s}\cdot\bm{p})\right] + \frac{\kappa}{2m_Q^2(E+m_Q)^2}\frac{\partial^2}{\partial s_i\partial s_j}\left[P\,s^2\,p_i\,p_j-2P(\bm{s} \cdot\bm{p})\,p_i\,s_j+P\,\delta_{ij}(\bm{s}\cdot\bm{p})^2\right]. \nonumber
\end{align} 
Because the noise is multiplicative, the Langevin equations in Eq.~\eqref{LE} are understood in the Stratonovich sense. The corresponding Fokker-Planck equation contains the associated drift correction, as shown in Eq.~\eqref{FP_gen}. This prescription also preserves the fixed magnitude of the spin vector under stochastic evolution.
\end{widetext}
%


\section{Spin diffusion and relaxation}

To obtain the Fokker-Planck equation for spin, we integrate over the momenta. We assume that the relaxation of momentum d.o.f happens much more quickly when compared to the spin d.o.f, i.e. $\tau_s\gg\tau_p$. This assumption can be justified because the dynamical participation of the spin d.o.f. occurs at a higher mass order~\cite{Manohar:2000dt} compared with the momentum diffusion case, which results in a longer relaxation time for the spin than for the momentum, by a factor of at least $m_{Q}/T$. Therefore, the momentum d.o.f. reaches equilibrium much earlier than the spin d.o.f. This adiabatic assumption allows us to write
\begin{align}\label{P}
P(\bm{p},\bm{s},t)=e^{-\beta E}f(\bm{s},t).
\end{align}
where $f(\bm{s},t)$ is the probability density of the spin vector of HQ to be aligned along $\vec{s}$ at time $t$. Employing this in Eq.~\eqref{FPps}, and integrating over momenta gives the Fokker-Planck equation for the spin of the HQ as
\begin{align}\label{spinFP}
\frac{\partial f}{\partial t}=\frac{2\kappa}{3}\frac{I_2}{I_1}\frac{\partial}{\partial s_i}\big(fs_i\big)+\frac{\kappa}{3}\frac{I_2}{I_1}\frac{\partial^2}{\partial s_i \partial s_j}\Big[f\left(\delta_{ij}s^2-s_is_j\right)\Big],
\end{align}
where 
\begin{align}
I_1=\int_0^{\infty}dp\,p^2e^{-\beta E} \,\,\, \text{and} \,\,\, I_2=\int_0^{\infty}dp\,\frac{p^4e^{-\beta E}}{m_Q^2(E+m_Q)^2}.
\end{align}
Here we have taken $\kappa$ to be independent of momentum. Interestingly, Eq.~\eqref{spinFP} has the same form as obtained in Ref.~\cite{Dey:2025ail}. From Eq.~\eqref{spinFP}, we can identify the spin drag and diffusion coefficients as
\begin{align}
A_i&=-\frac{2\kappa}{3}\frac{I_2}{I_1}s_i, \\
D_{ij}&=\frac{\kappa}{3}\frac{I_2}{I_1}(\delta_{ij}s^2-s_is_j)=\mathcal{D}(\delta_{ij}s^2-s_is_j).
\end{align}
Following Ref.~\cite{Dey:2025ail}, we define the spin relaxation time associated with Thomas precession of HQ as  
\begin{align}\label{tau_s}
\tau_s\equiv\frac{1}{\mathcal{D}}=\frac{3I_1}{\kappa I_2}. 
\end{align}
Ref.~\cite{Dey:2025ail} noted that $\tau_s$ should be related to dissipation in the medium through an Einstein-Stokes-like relation; Eq.~\eqref{tau_s} provides a relation of this kind for the contribution of the random momentum kicks, expressing $\tau_s$ in terms of momentum diffusion coefficient $\kappa$.

For a non-relativistic particle, using $\kappa=2\eta_Dm_QT$ \cite{Moore:2004tg}, up to leading order,
\begin{align}\label{tau_NR}
\tau_s^{NR}=\frac{4 m_Q^3}{\kappa T}=\frac{2m_Q^2}{\eta_DT^2}=2\Big(\frac{m_Q}{T}\Big)^2\tau_p.
\end{align}
where $\tau_p=1/\eta_D$. The resulting hierarchy $\tau_s/\tau_p=2(m_Q/T)^2\gg1$ demonstrates a posteriori that the momentum degrees of freedom can consistently be adiabatically eliminated in the heavy-quark limit, thereby justifying the assumption in Eq.~\eqref{P}. It is important to note that the above derivation is formulated for a homogeneous medium at fixed temperature. In an evolving QGP, the resulting relation may be applied locally with $T=T(t)$ and $\kappa=\kappa[T(t)]$.

Choosing the initial magnetic field to point along the $z$-axis, the spin Fokker-Planck Eq.~\eqref{spinFP} takes the form \cite{debye1929polar, kubo1970brownian}
\begin{align}\label{spinFPc}
\frac{\partial f}{\partial t}=\mathcal{D}\left[\frac{1}{\sin\theta}\frac{\partial}{\partial \theta}\Big(\sin\theta\frac{\partial f}{\partial \theta}\Big)+\frac{1}{\sin^2\theta}\frac{\partial^2f}{\partial \phi^2}\right].
\end{align}
Note that there are no derivatives over $|\bm{s}|$, which means that $f$ is a function only of $\theta$, $\phi$, and $t$ and it describes the probability density of a spin vector of fixed magnitude having an orientation ($\theta,\phi$) at time $t$. If the initial polarization of the HQ is along $\theta=\theta_0$, the initial distribution is given by $f(\theta,0)=\frac{1}{2\pi}\delta(\cos\theta -\cos\theta_0)$ \cite{Dey:2025ail}. With this choice of initial condition, the solution to Eq.~\eqref{spinFPc} is given as
\begin{align}
f(\theta,t)=\sum_{l=0}^{\infty}\frac{2l+1}{4\pi}e^{-l(l+1)t/\tau_s}P_l(\cos\theta)P_l(\cos\theta_0).
\end{align}
The vector polarization of the HQ along the direction of the initial field is given by the mean value of $\cos\theta$. In our case,
\begin{align}\label{avgcos}
\langle\cos\theta\rangle(t)=\cos\theta_0 \,e^{-2t/\tau_s}
\end{align}
in agreement with Ref.~\cite{Dey:2025ail}. The vector polarization therefore decays on the time scale $\tau_s/2$, while the tensor polarization, i.e.\ the deviation of $\langle\cos^2\theta\rangle$ from its isotropic value $1/3$, decays on the time scale $\tau_s/6$ \cite{Dey:2025ail}.

\begin{figure*}[ht]
    \centering
    \begin{subfigure}{0.45\textwidth}
        \centering
        \includegraphics[width=\textwidth]{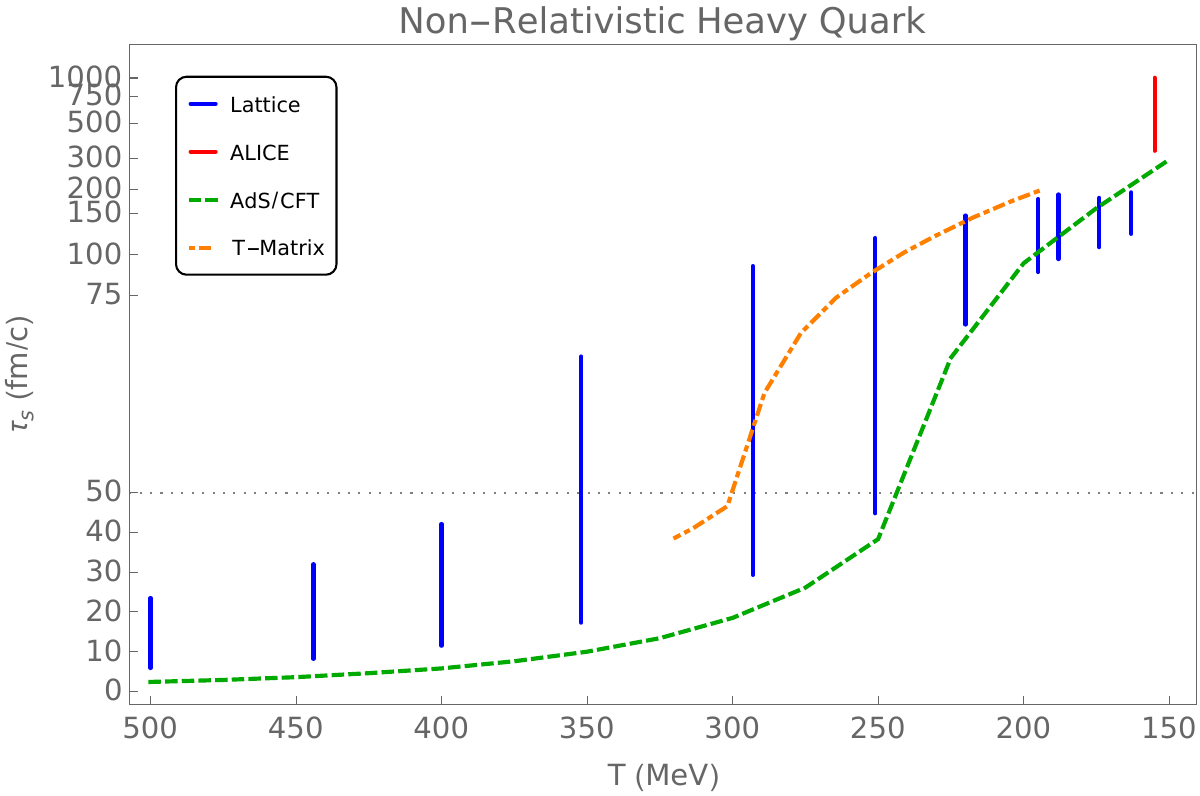} 
        \caption{\small }
        \label{dotted1}
    \end{subfigure}
    \begin{subfigure}{0.45\textwidth}
        \centering
        \includegraphics[width=\textwidth]{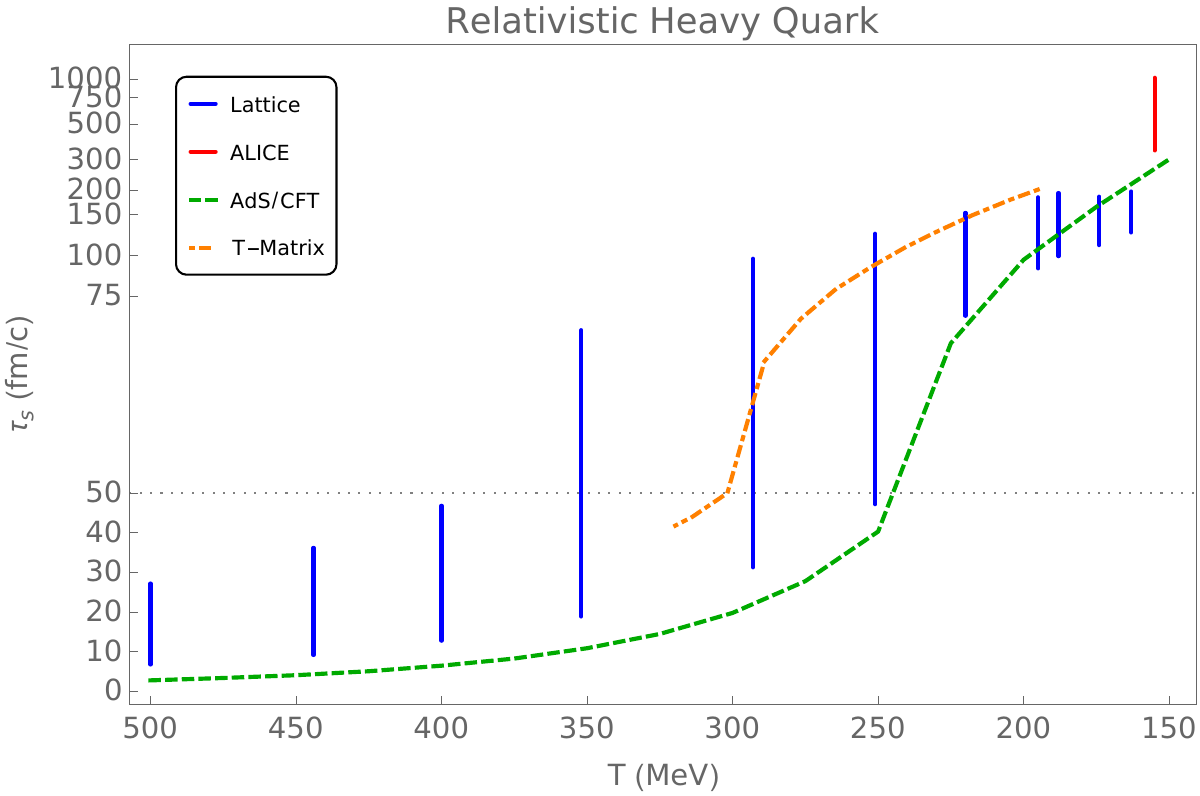} 
        \caption{\small }
        \label{dotted2}
    \end{subfigure}
\caption{\justifying \small Spin relaxation time $\tau_s$ of the HQ due to the random momentum kicks (Thomas precession), as a function of temperature, for different determinations of $\kappa$. Following Ref.~\cite{Dey:2025ail}, $\tau_s=1/\mathcal{D}$; the vector polarization of the HQ decays as $e^{-2t/\tau_s}$ (see Eq.~\eqref{avgcos}), i.e. on the time scale $\tau_s/2$. The Lattice results use $\kappa$ from Ref.~\cite{HotQCD:2025fbd}. For AdS/CFT ($\alpha_{SYM}=0.3$), ALICE, and T-matrix results, we used Refs.~\cite{PhysRevD.74.085012}, \cite{ALICE:2021rxa}, and \cite{Tang:2023tkm} respectively. The dotted line at $\tau_s=50\,\,\text{fm/c}$ is the demarcation for a change from linear to logarithmic scaling of the y-axis as one goes up in $\tau_s$ in both plots. The bars on the Lattice data are the uncertainties in the evaluation of $\kappa$.}
\label{fig1}
\end{figure*}

\section{Numerical estimates}

Fig.~\ref{fig1} shows the Thomas precession spin relaxation time $\tau_s$ of the charm quark as a function of temperature for the non-relativistic and relativistic dispersion relations in panels (a) and (b), respectively, using $\kappa$ obtained from four different approaches. The sources fall into two categories. The first comprises references where direct evaluation of $\kappa$ is given ~\cite{HotQCD:2025fbd, PhysRevD.74.085012, casalderrey2014gauge, GUBSER2008175, Casalderrey-Solana:2007ahi}, and the second category where the spatial diffusion coefficient $D_s$~\cite{Tang:2023tkm, ALICE:2021rxa} is given; for the latter, we recover $\kappa$ via the Einstein relation $\kappa=2T^2/D_s$ in the non-relativistic limit. Throughout we take $m_Q=1.27\,\,\text{GeV}$ \cite{ParticleDataGroup:2024cfk}. Here, we would like to emphasize that expressing $\tau_{s}$ in terms of $\kappa$ provides a very convenient approach for both the non-relativistic and relativistic cases, because in recent experimental data the spin observables of heavy hadrons are spanned over the momentum space up to $p_{T}\sim30\,m_{Q}$ \cite{ALICE:2025cdf}, where one needs to incorporate relativistic corrections.

In all cases, $\tau_s$ grows rapidly as the temperature falls. The strength of this growth, however, differs between methods and is controlled by the temperature dependence of $\kappa$. At any given temperature and $\kappa$, the relativistic and non-relativistic results differ only through the dispersion relation entering $I_1$ and $I_2$. The relative difference between $\tau_s^{R}$ and $\tau_s^{NR}$ is $\sim2.5\%$ near $T_c$, and rises to $\sim15\%$ at $T\sim500\,\,\text{MeV}$. Relativistic kinematics, therefore, have only a marginal effect on this contribution to HQ spin relaxation associated with Thomas precession, over the temperature range relevant to HICs. The relativistic estimates presented here incorporate the relativistic dispersion relation in the spin-kinematic factor, while retaining the momentum-independent isotropic diffusion kernel of Eq.~\eqref{variance}; a fully relativistic Langevin treatment with momentum-dependent longitudinal and transverse diffusion coefficients is beyond the scope of the present work.

The AdS/CFT estimate, corresponding to the strong-coupling limit, yields the shortest relaxation times for $T\gtrsim 220$~MeV, falling from $\sim295\,\,\text{fm/c}$ at $T\sim150\,\,\text{MeV}$ to $\sim2.4\,\,\text{fm/c}$ at $T\sim500\,\,\text{MeV}$ in the non-relativistic case. For the relativistic case, it ranges from $\sim302\,\,\text{fm/c}$ at $T\sim150\,\,\text{MeV}$ to $\sim2.8\,\,\text{fm/c}$ at $T\sim500\,\,\text{MeV}$. The lattice band was evaluated at eleven temperature points, spanning $163$ and $500\,\,\text{MeV}$, and its vertical width reflects the uncertainty in the lattice evaluation of $\kappa$. Since $\kappa/T^{3}$ is temperature independent in AdS/CFT, $\tau^{NR}_{s}$ scales with temperature and coupling strength $\lambda$ as $\tau^{NR}_s\propto \frac{1}{\sqrt{\lambda}T^{4}}$. 

The lattice $\kappa/T^{3}$ instead increases towards $T_c$, which weakens the growth of $\tau_s$. As a result, in both panels the AdS/CFT estimate lies below the lattice band above $T\sim220$~MeV, enters it between $195$ and $220$~MeV, and lies above it at $163$~MeV, in line with the observation that the lattice diffusion coefficient approaches the strong-coupling estimate near the crossover~\cite{HotQCD:2025fbd}.
The T-matrix result is shown over the range $195-320\,\,\text{MeV}$, in which that calculation is available, and carries a systematic uncertainty of $20-30\%$ quoted for the T-matrix transport coefficients~\cite{Tang:2023tkm}. The band labelled ALICE shows the range $1.5<2\pi D_sT_c<4.5$ at the pseudocritical temperature $T=155\,\,\text{MeV}$, spanned by the transport models that simultaneously describe the measured $R_{AA}$, $v_2$ and $v_3$ of the prompt D mesons in Pb-Pb collisions at $\sqrt{s_{NN}} = 5.02\,\,\text{TeV}$~\cite{ALICE:2021rxa,ALICE:2020iug}. It is quoted at a single temperature and is inferred from a model comparison rather than being a direct determination of $\kappa$.

For bottom quarks, this contribution is smaller still. Since $\tau_s^{NR}=4m^3_Q/(\kappa T)$, at equal $\kappa$ the bottom-quark spin relaxes $(m_b/m_c)^3\approx36$ times more slowly than that of charm for $m_b=4.18\,\,\text{GeV}$ \cite{ParticleDataGroup:2024cfk}; the lattice gives $\kappa_b$ slightly below $\kappa_c$ \cite{HotQCD:2025fbd}, which lengthens $\tau_s$ further.


\section{Direct chromomagnetic contribution}

The spin of the HQ can also relax through its direct coupling to the colour-magnetic fields of the medium, which is not included in the present treatment. The corresponding relaxation rate of the spin density has been computed in perturbative QCD at leading-log order \cite{Hongo:2022izs},
\begin{equation}
\Gamma_s^{LL} = C_F\,m_D^2\left[\frac{T\,g^2\ln(1/g)}{6\pi m_Q^2}\right].
\end{equation}
At the same order, the momentum diffusion coefficient and Debye mass are given by \cite{Moore:2004tg, HotQCD:2025fbd, Laine:2016hma}
\begin{equation}
\kappa = C_F\,D_F\frac{T^3\,g^4\ln(1/g)}{18\pi}, \quad m_D^2 = D_F\,\frac{g^2T^2}{3},
\end{equation}
where $D_F\equiv (N_c+N_f/2)$. Using these results, one finds
\begin{equation}
\Gamma_s^{LL} = \frac{\kappa}{m_Q^2} \implies \tau_s^{LL} \sim \frac{m_Q^2}{\kappa}.
\end{equation}
Comparing the above equation with the first equality in Eq.~\eqref{tau_NR}, we find that, at leading-log order, the spin-relaxation time associated with direct coupling to the chromomagnetic fields of the medium is smaller by a factor of order $m_Q/T$ than that associated with random Thomas precession, i.e., $\tau_s\sim(m_Q/T)\tau_s^{LL}$. 

The $\tau_s$ obtained in this work therefore accounts only for the contribution of the Thomas precession channel to the spin relaxation of HQs. This can be understood from the non-relativistic heavy-quark effective field theory~\cite{Braaten:1996ix}, where the direct coupling of the heavy-quark spin to the chromomagnetic field appears at order $1/m_Q$, whereas the chromoelectric spin-orbit interaction, which contains the Thomas-precession contribution, enters at order $1/m_Q^2$. Our findings give an effective estimate of the Thomas precession contribution by utilizing $\kappa$ values from the existing literature~\cite{HotQCD:2025fbd, PhysRevD.74.085012, casalderrey2014gauge, GUBSER2008175, Casalderrey-Solana:2007ahi} in Eq.~\eqref{tau_s}. The comparison with the direct chromomagnetic contribution is intended only as a leading-logarithmic, weak-coupling benchmark and should not be interpreted as a quantitative determination of the total spin-relaxation rate in the strongly coupled QGP. Our results motivate a systematic study of spin relaxation induced by the chromomagnetic and chromoelectric fields of the medium, which we leave for future work.

Moreover, the Thomas precession contribution to the spin diffusion coefficient is related to the momentum diffusion coefficient $\kappa$, so a more concrete understanding or estimation of the spin diffusion coefficient from theoretical and phenomenological perspectives can also provide another handle to constrain the value of $2\pi T D_{s}$. From our effective estimation, we have shown that the estimate of $\tau_{s}$ is only marginally affected in going from the non-relativistic to the relativistic case. However, one must keep in mind that the relativistic estimates presented here incorporate only the relativistic dispersion relation in the spin-kinematic factor while keeping a non-relativistic Langevin treatment. On the other hand, relativistic extensions of heavy-quark diffusion in the QGP have been investigated previously~\cite{Akamatsu:2008ge}; incorporating the corresponding relativistic corrections into the present framework is left for future work.


\section{Conclusion}

In this work, we have investigated the depolarization of heavy quarks in the QGP due to the random momentum kicks they receive from the medium, by constructing a Langevin equation for the heavy-quark spin from the classical Thomas-BMT equation, in which the same stochastic noise drives the diffusion of both momentum and spin. From it, we derived a spin-momentum Fokker-Planck equation in the extended phase space of the momentum and spin and, assuming that the momentum degrees of freedom equilibrate much faster than the spin, integrated over momenta to obtain the Fokker-Planck equation for the spin alone. This equation describes rotational diffusion on the spin sphere, with a spin relaxation time $\tau_s=3I_1/(\kappa I_2)$, which is fixed by the momentum diffusion coefficient $\kappa$ through a Thomas precession mechanism. The vector polarization of the HQ decays on the time scale $\tau_s/2$. In the non-relativistic limit, $\tau_s=4m_Q^3/\kappa T=2(m_Q/T)^2\tau_p$.

Using $\kappa$ from lattice QCD, AdS/CFT, the T-matrix approach, and transport-model constraints from D-meson observables, we evaluate $\tau_s$ for charm quarks over the temperature range $150$--$500$ MeV. At fixed $\kappa$, relativistic kinematics modify $\tau_s$ by only $2.5$--$15\%$ across this temperature range, indicating that the relativistic corrections are marginal. The AdS/CFT estimate yields the shortest relaxation times at high temperatures, while approaching the lattice band near the crossover, where the lattice $\kappa/T^3$ increases. Since $\tau_s\propto m_Q^3$, the Thomas-precession contribution is substantially weaker for bottom quarks, for which the relaxation time is about $36$ times larger than for charm at the same $\kappa$. Moreover, at leading-log order in perturbative QCD, direct coupling of the heavy-quark spin to the chromomagnetic fields~\cite{Hongo:2022izs} gives a relaxation rate larger by a factor of order $m_Q/T$ at fixed $\kappa$. This hierarchy reflects the fact that the present calculation captures only the Thomas-precession contribution induced by stochastic momentum kicks, and therefore represents just one channel of heavy-quark spin relaxation. Finally, If the different relaxation channels can be treated as independent and additive, a measured heavy-quark polarization decay rate can, in turn, be used to place an upper bound on $\kappa$.

At this juncture, we emphasize that the present analysis assumes a static and homogeneous medium. A realistic description of heavy-quark spin relaxation in an evolving QGP will require incorporating the space-time dependence of the medium, including its temperature evolution and flow. In an anisotropic medium, the stochastic momentum fluctuations may also become anisotropic, potentially modifying the spin-diffusion dynamics and giving rise to characteristic effects in polarization observables such as the polarization harmonics~\cite{Jaiswal:2026ixt}. Investigating these effects could provide a useful probe of the underlying medium dynamics and initial-state geometry. An important extension would also be to formulate the present framework with a fully relativistic heavy-quark diffusion kernel, including momentum-dependent longitudinal and transverse diffusion coefficients~\cite{Rajagopal:2026urq, Akamatsu:2008ge}. This is particularly relevant for heavy quarks at intermediate and high transverse momentum, where relativistic transport effects become important. Expressing the spin-relaxation time directly in terms of the momentum-diffusion coefficient $\kappa$, rather than the heavy-quark spatial diffusion coefficient $D_s$, makes the present framework more directly applicable to such transport calculations. Nevertheless, the estimates obtained here should be regarded as effective estimates of the Thomas-precession contribution, and a quantitatively complete treatment incorporating the full relativistic diffusion kernel and the space-time evolution of the QGP is left for future work.

The framework developed here may also find applications beyond spin physics in heavy-ion collisions, providing an avenue for studying non-equilibrium impurity-spin dynamics in quantum many-body systems~\cite{Lucas:2017idv, Akamatsu:2023yus}. In particular, comparing our result with the AdS/CFT estimate discussed above, one finds that $\tau_s\propto1/\sqrt{\lambda}$, indicating faster spin relaxation with increasing coupling strength in the holographic description. It would therefore be interesting to investigate the corresponding rotational Brownian motion in a suitable holographic model and explore how spin relaxation emerges in strongly coupled quantum many-body systems. More generally, extending the present approach to complex and evolving quantum environments could provide effective kinetic descriptions of impurity spin relaxation and transport and establish connections between stochastic spin dynamics and non-equilibrium phenomena across a broader class of quantum many-body systems. These extensions constitute interesting directions for future work.

\bibliography{mainref}

\end{document}